\documentstyle{article}

\normalsize

\begin{document}

\bibliographystyle{unsrt}
\footskip 1.0cm
\thispagestyle{empty}
\setcounter{page}{0}
\begin{flushright}
IC/94/224\\

August 1994\\
\end{flushright}
\vspace{10mm}

\centerline {\large G. Bimonte $^{(1,2)}$}
\vspace{5mm}
\centerline{ \small and }
\vspace*{5mm}
\centerline{ \large G. Lozano $^{(1)}$ \footnote{
E-mail addresses: Bimonte@ictp.trieste.it~~,~~Lozano@ictp.trieste.it}}
\vspace*{5mm}
\centerline {\it (1) International Centre for Theoretical Physics, P.O.BOX
586} \centerline {\it I-34100 Trieste, ITALY}
\vspace*{5mm}
\centerline {\it (2) INFN, Sezione di Napoli, Napoli, ITALY}
\vspace*{25mm}
%\baselinestretch{2.0}
\normalsize
\centerline {\bf Abstract}
\vspace*{5mm}
{\large We consider an Electroweak string in the background of a uniform
distribution of cold fermionic matter. As a consequence of the fermion
number non-conservation in the Weinberg-Salam model, the string produces
a long-range magnetic field.} \newpage

\baselineskip=24pt
\setcounter{page}{1}

\newcommand{\beq}{\begin{equation}}
\newcommand{\eeq}{\end{equation}}

Since the pionering work by Nielsen and Olesen~\cite{NO}, the study of
vortex like
solutions in Gauge Theories has attracted a constant interest. This kind
of objects is by now believed to play a prominent role in cosmology, for
example by providing a mechanism to explain structure formation in the
early universe~\cite{B}.

Until very recently, this type of configurations have been mainly studied
in theories with a non trivial vacuum structure where the existence of a
topological conservation law ensures its stability. However, as
recently  stressed  by Vachaspati and Barriola~\cite{VB}, the embedding of
such configurations in
larger  gauge groups can be stable even in the absence of such topological
conservation laws. This is what happens in the case of the
Z-string~\cite{V}, which results from
embedding  the U(1) Nielsen Olesen string in the $SU(2)\times U(1)$
theory of Electroweak interactions~\cite{NMES}.

In order to understand the role of these configurations during the phase
transition it is important to know how they are modified in typical
thermodynamical situations such as non-zero temperature and finite matter
density. In this letter we shall analyze the case in which the string is
placed in a background of cold neutral fermionic matter,
 a situation in which, due to the V-A character of the Electroweak
interactions, the fermion sector of the model manifests in a non trivial way.

To illustrate the kind of phenomena that occur, let us begin with the
simpler $U(1)$ model and study the way in which a constant fermionic density
affects the Nielsen Olesen string. We consider the system described by
the Lagrangian density,
\beq
L=-\frac{1}{4}Z_{\mu\nu}Z^{\mu\nu} + (D_\mu\phi)(D^\mu\phi) -
\lambda(\phi^2-\eta^2)^2 + L_f \,\,\,.
\eeq
Here the covariant derivative for the scalar complex field $\phi$ is
defined as
\beq
D_{\mu}\phi = (\partial_{\mu} + i q Z_{\mu}) \phi
\eeq
and $Z_{\mu}$ and $Z_{\mu\nu}$ are the vector potential and field strength.

The interaction of the fermions with the gauge fields is governed by the
Lagrangian density
\beq
L_f =\bar{\psi}_L^-  i \sigma_{\mu} ( \partial_\mu -i\frac{q}{2}
Z_{\mu})\psi_L^-
+ \bar{\psi}_L^+ i \sigma_{\mu} ( \partial_\mu +i\frac{q}{2} Z_{\mu} )
\psi_L^+ \eeq
where $\psi_L^-$ and $\psi_L^+$ are
 left handed fermions with opposite charge and $\sigma_{\mu}$
are the Pauli matrices.
Following Rubakov and Tavkhelidze~\cite{RT},
we will consider the case in which there is a cold (zero temperature)
and neutral ($ n^+=n^-=n/2 $) background of fermions.

In the weakly coupled regime, the
bosonic field can be treated
classically and its contribution to the energy (in the static case) is
simply given by
\beq
E_b=\int d^3x [ \frac{1}{4} Z_{ij}^2 + |D_i\phi|^2 + \lambda
(\phi^2-\eta^2)^2] ~~.
\eeq
As shown in Refs.~\cite{RT},~\cite{RWT} the presence of a finite density
manifests through the appearance in the  effective energy of a Chern Simons
term~\cite{JTS} which results from the integration of the fermionic
determinant
 \beq
E_f= -\frac{k}{2} \int d^3x \epsilon^{ijk}Z_{ij} Z_k
\eeq
the coefficient $k$ being  related to the chemical potential $\mu$ and
fermion density $n$ by
\beq
k= \frac{\mu q^2}{16\pi^2} = \frac{q^2}{16\pi^2} (3\pi^2n)^{\frac{1}{3}}~.
\eeq

The effective energy
\beq
E=E_b +E_f
\eeq
was originally used in Refs ~\cite{RT} ~\cite{RMRT}
where it was shown that the trivial vacuum becomes unstable for chemical
potentials larger
than the critical value
\beq
\mu_{crit} = \frac{32\pi^2}{q^2} m_v
\eeq
where the mass of the gauge particle is $m_v^2=2 q^2 \eta^2$.
We will instead consider the case of small chemical potentials
$\mu \leq \mu_{crit}$ in order to study how the fermionic background
modifies the Nielsen Olesen string.

Notice that by making in (7) the transformation
\beq
A_3 \rightarrow iA_0 \,\,\,
x^3 \rightarrow  -ix^0 \,\,\,
k \rightarrow  ik
\eeq
we obtain the {\it action} of the $2+1$ dimensional Chern-Simons-Higgs
system. Vortex like solutions in this and related models have been
object of much study in recent years mainly in the context of High
Temperature Superconductivity and the Quantum Hall
Effect~\cite{PKVS}~\cite{ILMS}. A general
feature of these solutions is that as a result of the Chern Simons
interaction the vortex acquires an electric charge and consequently a radial
electric field.

The transformation (9) corresponds to the combined effect of an Euclidean
rotation and a duality transformation in the $x_0-x_3$ plane. As a
consequence, what in the $2+1$ dimensional case corresponded to an
electric charge will now  become a flux of current along the string and
the radial electric field will be now converted into a tangential magnetic
field. Making the ansatz
\beq
\phi= f(r) \exp(i m\theta) \,\,\,\,\,\,\,\, {\bf Z}=Z_{\theta}{\bf
e_{\theta}} + Z_3 {\bf e_3}
\eeq
the equations of motion then read
\begin{eqnarray}
 (rZ_3')' -2q^2rf^2Z_3 + 2kZ'_{\theta} = 0 \\
\left(\frac{Z'_\theta}{r}\right)'-2q \frac{m+qZ_{\theta}}{r}f^2-2kZ_3'= 0 \\
\frac{1}{r}(rf')'-\frac{(m + qZ_{\theta})^2}{r^2} f -q^2 Z_3^2f-2\lambda f
(f^2-\eta^2)=0 \end{eqnarray}
subject to the boundary conditions
\begin{eqnarray}
Z_3(\infty)=0 & Z_{\theta}(\infty)= -\frac{m}{q} & f(\infty) =\eta \\
Z_3(0) < \infty & Z_{\theta}(0)=0 & f(0)=0~~.
\end{eqnarray}

Although an exact solution to the equations cannot be obtained
analytically, it is possible ( in complete analogy to the $2+1$ dimensional
case ~\cite{PKVS}~\cite{ILMS}) to express
the first terms of an expansion in $k$ in a closed form involving the
solutions of the Nielsen Olesen vortex
\begin{eqnarray}
Z_{\theta}=Z_{\theta}^{NO} -\frac{k^2 r^2}{2}(Z_{\theta}^{NO}+\frac{m}{q})+
O(k^4) \\ Z_3=-k(Z_{\theta}^{NO}+ \frac{m}{q}) + O(k^3) \\
f=f^{NO} + O(k^4)
\end{eqnarray}
where the superscript $NO$ refers to the Nielsen Olesen solutions, that
is, solutions to equations (11-13) with $k=0$.
Thus, we see that the leading effect is the appearance of a tangential
component of the {\it quasi magnetic} field $C_i$ (we reserve the name
magnetic field for later when we introduce the unbroken U(1) of
electromagnetism)
\begin{eqnarray}
C_i&=&\frac{1}{2} \epsilon_{ijk}Z_{jk} \\
C_{\theta}&=&-rZ_3'
\end{eqnarray}
which can be associated to a flux of current flowing along the string
\beq
J = \int d^2x j_3 \,\,\,\,\,\,\,\,
j_l= iq(\phi D_l\phi^* - \phi^*D_l\phi)~~.
\eeq
Using the boundary conditions, it can be easily shown that
\beq
J=2k \Phi~~,
\eeq
where $\Phi$ is the flux of the quasi-magnetic field
\beq
\Phi=\int d^2x C_3=-2\pi \frac{m}{q}~~.
\eeq
The asymptotic behavior for the gauge fields at infinity can be derived from
(11-13)
\begin{eqnarray}
Z_3 &\rightarrow & \frac{ic}{\sqrt{r}} e^{-\mu r} + h.c. \\
Z_{\theta} &\rightarrow & -\frac{m}{q} + (c \sqrt{r} e^{-\mu r} + h.c.)
\end{eqnarray}
where
\beq
\mu = \sqrt{(m_v^2-k^2)} -ik  \,\,\,\,\,\,\, m_v^2=2q^2\eta^2~~.
\eeq
Comparing this result with the $2+1$ dimensional case, we see that as a
consequence of the replacement $k\rightarrow ik$, $\mu$ has became a
complex number. Notice that as far as
$
k^2\leq m_v^2
$
or equivantly
$
\mu\leq\mu_{crit}
$,
$\mu$ has a positive real part. In addition, this real part is smaller
compared to the case $k=0$ which clearly corresponds to the anti-screening
effect of the Chern Simons term. We then see, that the value $\mu=\mu_{crit}$
corresponds to the point for which this effect completely overwhelms the
screening originating from the Higgs mechanism.

As for the Higgs field, its asymptotic is
\begin{eqnarray}
f &\rightarrow & \frac{D_1}{r} e^{-\mu r} +
\frac{D_2}{\sqrt{r}} e^{-m_H r} + h.c. \\
 m_H^2 &=& 4\lambda \eta^2
\end{eqnarray}
and depending on the ratio $\frac{|\mu|}{m_H}$ the behavior at
infinity will be determined by the first or second term~\cite{P}.

We can now turn to the study of the Electroweak case. The contribution of
the bosons to the static effective energy is
\beq
E_b=\int d^3x [ \frac{1}{4}(W_{ij}^a)^2 + \frac{1}{4} B_{ij}^2 +
|D_i\Phi|^2 + \lambda (\Phi^2-\eta^2)^2]
\eeq
where the covariant derivative of the Higgs doublet and the  strengths
of the $SU(2)\times U(1)$ gauge fields are defined as
\begin{eqnarray}
D_i\Phi&=& (\partial_i - i\frac{g}{2} \tau^a W_i^a -
i\frac{g'}{2}B_i)\Phi          \\
W_{ij}^a &=&\partial_iW_j^a-\partial_jW_i^a + g \epsilon^{abc}W_i^aW_j^b \\
B_{ij}&=&\partial_iB_j-\partial_jB_i~~.
\end{eqnarray}

Assuming a cold background of fermions satisfying
\beq
n_{e_L}^{(i)}
=
n_{\nu}^{(i)}
=
n_{d_L}^{(i) a }
=
n_{u_L}^{(i) a}
=
n_{u_R}^{(i) a }
=
n_{d_R}^{(i) a }
=
n_{e_R}^{(i)}
\eeq
(where $i=1,..,f$ and $a=1,..3$ denote the family and color indices
respectively) their contribution to the effective energy is given
by~\cite{RT},~\cite{RWT}:
\begin{eqnarray}
E_F&=& -4f\mu[ N_{CS}(W)-N_{CS}(B) ] \\
N_{CS}(W)&=& \frac{g^2}{32\pi^2}\int d^3x
\epsilon^{ijk}(W_{ij}^aW_k^a-\frac{g}{3}\epsilon^{abc}W_i^aW_j^bW_k^c) \\
N_{CS}(B)&=&\frac{g'^2}{32\pi^2}\int d^3x\epsilon^{ijk}B_{ij}B_k
\end{eqnarray}

The effects of the Chern Simons term on the ground state of the Weinberg
Salam model in an external magnetic field and its influence on the
phenomenon of W-condensation~\cite{AO} has been recently analized by
Poppitz~\cite{P}.
We will instead concentrate here on the influence of this term on the
Z-string configuration and consequently  we
will set to zero
the charged W bosons and the upper component of the Higgs field. Denoting
as $\phi$ the lower component of the Higgs,  the energy
takes the form
\beq
E=\int d^3x \left[ \frac{1}{4}(F_{ij})^2 + \frac{1}{4} Z_{ij}^2 +
|D_i\phi|^2 + \lambda (\phi^2-\eta^2)^2 - \epsilon^{ijk}(k_1 Z_iF_{jk}
+\frac{k_2}{2}Z_iZ_{jk}) \right]~~,
\eeq
where we have introduced the electromagnetic vector potential $A_i$ and
the neutral vector potential $Z_i$ by
\begin{eqnarray}
A_i&=&W^3_i \sin\theta  + B_i \cos\theta  \\
Z_i&=&W^3_i \cos\theta  - B_i \sin\theta  \\
\tan\theta &=&\frac{g'}{g}
\end{eqnarray}
and $k_1$ and $k_2$ are defined as
\beq
k_1=\frac{g g'}{4\pi^2} \mu f \;\;\;\,\,\,\, k_2=\frac{g^2-g'^2}{4
\pi^2}\mu f~~.
\eeq

Notice the absence of Chern Simons term associated to the field
 $A_{i}$, a fact
which can be traced back to the vector character of the electromagnetic
interactions. Due to the presence of the cross Chern Simons term, the
field $Z_{i}$ will now act as a source for $A_i$ and it is not any more
consistent (as in the purely bosonic case~\cite{V}) to set it to zero.
Indeed, the equations for $A_i$ are
\beq
\partial_iF^{il} = - k_1 \epsilon^{kjl} Z_{kj}~~.
\eeq
For $\mu$ (i.e. $k$) small compared with the particle masses, we can
neglect the back reaction of the condensate on the string. Then,
replacing the right hand side of (42) by its value in the Nielsen Olesen
case, we obtain a tangential magnetic field
\beq
{\bf H}= 2 k_1 \frac{Z_{\theta}^{NO}}{r} {\bf e_{\theta}}~~.
\eeq
At, large distances, where $Z_{\theta}^{NO}$ goes to
$-m/q$ ($q=\frac{\sqrt{g^2+g'^2}}{2}$)
\beq
{\bf H}= -\frac{em \mu f}{\pi^2 r} {\bf e_{\theta}}
\eeq
which coincides with the magnetic field generated by an infinitely long
wire carrying a current
\beq
I=-\frac{em\mu f}{2\pi^2}~~.
\eeq
As in the abelian case, there will be  a tangential component of
the quasi-magnetic field
\beq
{\bf C}=-r Z_3'{\bf e_{\theta}}=-k_2 r (Z_{\theta}^{NO})' {\bf e_\theta}~~.
\eeq
We can estimate the maximum value of the magnetic
field to be
\beq
H_{max}\sim \frac{k_1 m_v m}{q}~~. \;\;\;\;
\eeq

As we mentioned before, all these considerations are valid for small
chemical potentials and strictly speaking for small distances from the core
of the string.

A more complete analysis would follow from the numerical study of the
equations of motion which result from the ansatz

\beq
\phi= f(r) \exp(i m\theta) \,\,\,\,\,\,\, {\bf Z}=Z_{\theta}{\bf
e_{\theta}} +
Z_3 {\bf e_3} \,\,\,\,\,\,\,  {\bf A}= A_{\theta}(r) {\bf e_{\theta}} +
A_3(r) {\bf e_3}~~,
\eeq
\begin{eqnarray}
 (rZ_3')' -2q^2rf^2Z_3 + 2k_2Z'_{\theta} +4k_1^2rZ_3= 0 \\
\left(\frac{Z'_\theta}{r}\right)'-2q
\frac{m+qZ_{\theta}}{r}f^2-2k_2Z_3'+4k_1^2\frac{Z_\theta}{r} = 0 \\
\frac{1}{r}(rf')'-\frac{(m + qZ_{\theta})^2}{r^2} f -q^2 Z_3^2f-2\lambda f
(f^2-\eta^2)=0~~.
\end{eqnarray}
Once the solutions are known, the electromagnetic potentials are
calculated by direct integration (see eqs (42)).
At infinity, due to the long range character of electromagnetism, we expect
a substantial modification of the Z-string profile. Indeed, the
asymptotic behavior
of the massive fields will turn from exponential to power like and
additionally there will be a modification of the total Z-flux. To the
leading order in the chemical potential we obtain
\begin{eqnarray}
Z_{\theta} &\rightarrow &-\frac{m}{q} -k^2_1c_1 \\
Z_3 &\rightarrow & -\frac{c_3 k_1^6 k_2}{r^4} \\
f  &\rightarrow & \eta -\frac{c_4 k_1^4}{r^2} \\
A_3 &\rightarrow & \frac{2m}{q}k_1 \ln(r) \\
A_\theta &\rightarrow & -k_1^7k_2 c_3 \frac{1}{r^2}~~.
\end{eqnarray}

Summarizing, we have shown that the presence of  cold neutral matter
considerably modifies the original Z-string  profile leading to the
appearance of long range magnetic fields associated with currents flowing
along the string.

Clearly, an important issue to be addressed concerns the stability of
these configurations. As it was established in Ref.~\cite{JPV}, the
bare Electroweak string is unstable for realistic values of the coupling
constants. The question is then to study whether the fermions can improve
the stability of the Z-string. Although the answer of this issue in the
full Electroweak model seems rather complicated, an encouraging
result comes from
the analysis of the semilocal string. In fact, for this case, stability is
improved. This can be shown by looking at the potential
term in the Schroedinger-like
equation satisfied by the perturbations and noticing that it is less
negative than in
the bare case. It would be interesting to explore in the Electroweak
theory the region of stability as a function of the coupling constants
and the fermion density (and eventually include temperature corrections)
in order to establish the role of these configurations in a cosmological
setting.

We would like to thank Prof. Abdus Salam, the International Atomic Energy
Agency and UNESCO for hospitality at the International Centre
for Theoretical Physics.

\end{document}